\documentclass[pre]{revtex4}
\usepackage{graphicx,color}
\usepackage{epstopdf}

\begin{document}

\title{The dynamics of human body weight change}

\author{Carson C. Chow\footnote{corresponding author: email:carsonc@mail.nih.gov} and Kevin D. Hall}
\affiliation{Laboratory of Biological Modeling, NIDDK. NIH, Bethesda, MD 20892}
\date{\today}

%\author{Carson C. Chow\thanks{To whom correspondence should be
%    addressed: Email:carsonc@mail.nih.gov} 
%\affil{1}{Laboratory of Biological Modeling,
% NIDDK, NIH, Bethesda, MD 20892}
%\and Kevin D. Hall\affil{1}{}}

%\maketitle

%\begin{article}
 
 \begin{abstract}

An imbalance between energy intake and energy expenditure will lead to
a change in body weight (mass) and body composition (fat and lean
masses).  A
quantitative understanding of the processes involved, which currently
remains lacking, will be useful in determining the
etiology and treatment of obesity and other conditions resulting from prolonged energy imbalance.
Here, we show that the long-term dynamics of human weight change can
be captured by  
a mathematical model of the macronutrient flux balances and all 
previous models are special cases of this model. We show that the
generic dynamical 
behavior of body composition for a clamped diet can be divided
into two classes.  In the
first class, the body composition and mass 
are determined uniquely. In
the second class, 
the body composition can exist at an infinite number of possible
states.
Surprisingly, perturbations of 
dietary energy intake or energy expenditure can give identical responses
in both model classes and  existing data are insufficient
to distinguish between these two possibilities.   However, this distinction is important for the efficacy of clinical interventions that alter body composition and mass. \\

\noindent
{\bf Summary}

Understanding the dynamics of human body weight change has important
consequences for conditions such as obesity, cancer, AIDS, anorexia
and bulimia nervosa. While we know that changes of body weight result
from imbalances between the energy derived from food and the energy
expended to maintain life and perform physical work, quantifying this
relationship has proved difficult. Part of the difficulty stems from
the fact that the body is comprised of multiple components and we 
must quantify how weight change is reflected in terms of alterations
of body composition (i.e. fat versus lean mass).  Mathematical modeling
is a means to  
address this issue.  Here we show that a mathematical model of the
flux balances of macronutrients, namely fat, protein and carbohydrates, can provide a general description
of  the way the body weight will change over time.  For a fixed food
intake rate and physical activity level, the body weight and body
composition will approach a steady state. However, the steady state can correspond to a unique body weight or a continuum of body weights depending on how fat oxidation depends on the body weight and composition changes.  Interestingly, the
existing experimental data on human body weight dynamics cannot
presently distinguish between these two possibilities.  We propose
experiments that could determine if steady state body weight is unique
and use computer simulations to demonstrate how such experiments could
be performed.

 \end{abstract}
 
%\keywords{obesity | mathematical modeling | dietary intake | energy expenditure}

\maketitle

 \section{Introduction}

Obesity, anorexia nervosa, cachexia, and starvation are conditions
that have a
profound medical, social and economic impact on our lives.  For
example,  
the incidence of obesity and its co-morbidities has increased at a rapid rate over the past two
decades~\cite{hill2003,hill2006}.
These conditions
are characterized by changes in body weight (mass) that arise
from an imbalance between the energy derived  
from food and the energy expended to maintain life and perform work.
However, the underlying mechanisms of how changes in energy balance lead
to changes in body mass and body composition are not well understood.
In particular, it is of interest
to understand how body
composition is apportioned between fat and lean components when the
body mass changes and if this energy partitioning can be altered.  
Such an
understanding  would be useful in 
optimizing weight loss treatments in obese subjects to maximize fat loss or
weight gain treatments for anorexia nervosa and cachexia patients to maximize
lean tissue gain.

To address these issues and improve our understanding of human body
weight regulation, mathematical and computational modeling has been
attempted many times over the past several decades~\cite{hall2006,hall2002,alpert1979,alpert1990,alpert2005,Forbes1970,Livingston2001,hbc,jordan,antonetti1973,payne1977b,Girardier1994,westerterp1995,kozusko2001,flatt2004,christiansen2005,song2007}. 
Here we show how models of body composition and mass change can be understood
and analyzed within 
the realm of dynamical systems theory and
can be classified according to their geometric
structure in the two dimensional phase plane.  We begin by considering
a general class of macronutrient flux balance equations and
progressively introduce assumptions that constrain the
model dynamics. We show that two compartment models of fat and lean
masses can be categorized into 
two generic classes. In the first class,
there is a unique body composition and mass (i.e. stable fixed point)
that is specified by the diet and energy expenditure. In the
second class,  there is a continuous curve of
fixed points (i.e. invariant manifold) with an infinite number of
possible body compositions and masses at steady state for the same diet and energy expenditure.
We show that almost all  of the
models in the literature are in the second class.  Surprisingly, the
existing data are insufficient to
determine which of the two classes pertains to humans.
For models with an invariant manifold, we show that an
equivalent one dimensional equation for body composition change can be
derived.  We give numerical examples and discuss possible experimental
approaches that may distinguish between the classes.

\section{Results}
\subsection{General model of macronutrient and energy flux balance}
 
The human body obeys the law of energy conservation~\cite{atwater},
which can be expressed as
\begin{equation}
\Delta U= \Delta Q - \Delta W
\label{eq:cons}
\end{equation}
where $\Delta U$ is the change in stored energy in the body, $\Delta
Q$ is a change 
in energy input or intake, and $\Delta W$ is a change in energy output
or expenditure. 
The intake is provided by the energy content of the food
consumed.  
Combustion of dietary macronutrients yields chemical energy and Hess's
law states that the energy released is the same regardless of 
whether the process takes place inside a bomb calorimeter or via the
complex process of oxidative phosphorylation in the
mitochondria. Thus, the energy that can be derived from combustion of
food in the body can be precisely measured in the laboratory. However,
there is an important caveat. Not all macronutrients in food are
completely absorbed by the body. Furthermore, the dietary protein that
is absorbed does not undergo complete combustion in the body, but
rather produces urea and ammonia. In accounting for these effects, we
refer to the metabolizable energy content of dietary
carbohydrate, fat, and protein which are slightly less than the values
obtained by bomb calorimetry. 
The energy expenditure includes  the work to
maintain basic metabolic function (resting metabolic rate), to digest,
absorb and transport the nutrients in 
food (thermic effect of feeding), to synthesize or break down tissue,
and to perform physical activity, together with the heat generated.
The energy is stored in the
form of fat as well as in lean body tissue such as
glycogen and protein.  The body need not be in 
equilibrium for Eq.~(\ref{eq:cons}) to hold.  While we are primarily
concerned with adult weight change, Eq.~(\ref{eq:cons}) is also valid
for childhood growth.

In order to express a change of stored energy $\Delta U$ in terms of  body
mass $M$ we must determine
the energy content per unit body mass change, i.e.~the energy
density $\rho_M$.  We can then set $\Delta U = \Delta(\rho_M M)$. To model
the dynamics of body mass change, we divide
Eq.~(\ref{eq:cons}) by some interval of time and take the limit of
infinitesimal change to obtain a one dimensional energy flux balance
equation:  
\begin{equation}
 \frac{d}{dt}(\rho_M M)= I - E
\label{eq:M}
\end{equation}
where $I=dQ/dt$ is the rate of metabolizable energy intake and
$E=dW/dt$ is the rate 
of energy 
expenditure.  It is important to note that $\rho_M$ is the energy
density of body mass change, which
need not be a constant but could be a function of body composition and time.  Thus, in order to use Eq.~(\ref{eq:M}), the
dynamics of $\rho_M$ must also be established.    

When the body changes mass, that change will be composed of water,
protein, carbohydrates (in the form of glycogen), fat, bone, and trace
amounts of micronutrients,
all having their own energy densities.  Hence, a means of
determining the dynamics of $\rho_M$ is to track the dynamics of the 
components.  
The extracellular water and bone mineral mass have no metabolizable energy
content and change little when
body mass changes in adults under normal conditions~\cite{keys1950}.
The change in intracellular water
can be specified by changes in the
tissue protein and glycogen.  
Thus the main
components contributing to the dynamics of  $\rho_M$ are the
macronutrients - protein, 
carbohydrates, and fat, where we distinguish body fat
(e.g. free fatty acids and triglycerides) from adipose tissue, which
includes
water and protein in addition to triglycerides.  
We then represent Eq.~(\ref{eq:M}) in terms of
 macronutrient flux
balance equations for body fat $F$, glycogen $G$, and protein $P$:
\begin{eqnarray}
\rho_F\frac{dF}{dt}&=&I_F-f_F E\label{eq:ffat}\\
\rho_G\frac{dG}{dt}&=&I_C-f_C E\label{eq:glyc}\\
\rho_P\frac{dP}{dt}&=&I_P-(1-f_F-f_C) E \label{eq:prot}
\end{eqnarray}
where $\rho_F=39.5$ MJ/kg, $\rho_G=17.6$ MJ/kg, $\rho_P=19.7$ MJ/kg are the
energy densities, $I_F, I_C, I_P$ are the intake rates, and $f_F$, $f_C$,
 $1-f_F-f_C$ are the fractions of the energy expenditure rate obtained
from the combustion
of fat, carbohydrates (glycogen) and
protein respectively. The fractions and energy expenditure rate are
functions of body composition and intake rates.  They can be estimated from
indirect calorimetry, which measures the oxygen consumed and carbon
dioxide produced by a subject~\cite{atwater1905}.
The intake rates are determined by the macronutrient composition of
the consumed food, the efficiency
of the conversion of the food into utilizable form and the thermic
effect of feeding.  
Transfer between
compartments such as {\it de novo} lipogenesis where 
carbohydrates are
converted to fat or gluconeogenesis where amino acids
are converted into carbohydrates can be 
accounted for in the forms of $f_F$ and $f_C$.
The sum of  Eqs.~(\ref{eq:ffat}), (\ref{eq:glyc}) and (\ref{eq:prot})
 recovers the energy flux
balance equation~(\ref{eq:M}), where the body mass $M$ is the sum of the macronutrients $F$, $G$, $P$,  with the associated intracellular water, and the inert mass that does not change such as the extracellular water, bones, and minerals, and $\rho_M=(\rho_F F+\rho_G G +\rho_P P)/M$.

The intake and energy expenditure rates are explicit functions
of time with fast fluctuations on a time scale of hours to days
\cite{periwal2006}.  However, we are interested in the long term 
dynamics over weeks, months and years.  Hence, to simplify the equations, we
can use the method of averaging to remove the fast motion and
derive a system of equations for the slow time dynamics.   We do this
explicitly in the Methods section and show that the form of the averaged
equations to  lowest order are identical to
(\ref{eq:ffat})-(\ref{eq:prot}) except that 
the three components are to be interpreted as the slowly varying part
and  the intake and energy expenditure rates are moving time averages over a time scale of a day.

The three compartment flux balance model was used by
Hall~\cite{hall2006} to numerically simulate data from the 
classic Minnesota human starvation experiment \cite{keys1950}.  In
Hall's model, the  
forms of the energy expenditure and fractions were chosen for
physiological considerations.   For clamped food
intake, the body composition approached a unique steady state.
The model also showed that
apart from transient changes lasting only a few days, carbohydrate
balance is precisely maintained as a result of the limited storage
capacity for glycogen. We will exploit this property to
reduce the three dimensional system to an approximately equivalent two
dimensional system where dynamical systems techniques can be employed to
analyze the system dynamics.

\subsection{Reduced models}\label{mac}

\subsubsection{Two compartment macronutrient partition model}

The three compartment macronutrient flux balance model
Eqs.~(\ref{eq:ffat})-(\ref{eq:prot}) can be reduced
to a two 
dimensional system for fat mass $F$ and lean mass 
$L=M-F$, where $M$ is the total body mass.
The lean mass includes
the protein and glycogen with the associated intracellular water
along with the mass that does not change
appreciably such as the extracellular water and bone.  Hence the
rate of change in lean mass is given by
\begin{equation}
\frac{dL}{dt}=(1+h_P)\frac{dP}{dt} +(1+h_G)\frac{dG}{dt}
\label{leanmass}
\end{equation}
where $h_P=1.6$ and $h_G=2.7$ are reasonable estimates of the
hydration coefficients for the 
intracellular water associated with the
protein and glycogen respectively~\cite{hall2006,hall2007}.  (We note that fat is not associated with any water.) The 
glycogen storage capacity is extremely small compared to the
fat and protein compartments.  Thus the slow component of glycogen
can be considered to be a constant (see Methods).  In other words,
 on time scales much 
longer than a day, which are of interest for body weight change,
we can consider glycogen to be in quasi-equilibrium so that
$dG/dt=0$, as observed in numerical simulations~\cite{hall2006}.
This implies that  $f_C=I_C/E$, which can be substituted into
Eq.~(\ref{eq:prot}) to give
\begin{equation}
\rho_P\frac{dP}{dt}=I_P+I_C-(1-f_F) E
\label{PP}
\end{equation}
Substituting Eq. (\ref{PP}) and $dG/dt=0$ into Eq. (\ref{leanmass})
leads to
the {\it two compartment macronutrient partition} model 
  \begin{eqnarray}
\rho_F\frac{dF}{dt}&=&I_F-f E\label{eq:fat}\\
\rho_L\frac{dL}{dt}&=&I_L-(1-f)E\label{eq:lean}
\end{eqnarray}
where $\rho_L=\rho_P/(1+h_P)=7.6$ MJ/kg,
 $I_F$ and $I_L=I_P+I_C$ are the intake rates into the fat
and lean compartments respectively,  $E=E(I_F,I_L,F,L)$ is the
 total energy expenditure rate, and
$f= f(I_F,I_L,F,L)\equiv f_F$ is the fraction of energy expenditure rate
attributed to fat utilization.

We note that $dG/dt=0$ may be violated if the glycogen content is
proportional to the protein content, which is plausible because
the glycogen mass is stored in muscle tissue and may scale with
protein mass.  We show that this assumption leads to the same two
dimensional system. Substituting
\begin{equation}
\frac{dG}{dt} = k \frac{dP}{dt}
\label{newdg}
\end{equation}
for a proportionality constant $k$,  into
Eq.~(\ref{eq:glyc}) gives 
$f_CE=I_C-\rho_Ck dP/dt$
which inserted into Eq. (\ref{eq:prot}) leads to
\begin{equation}
(\rho_P+k\rho_C)\frac{dP}{dt}=I_L-(1-f)E
\label{PC}
\end{equation}
Substituting Eqs.  (\ref{newdg}) and (\ref{PC}) into Eq. (\ref{leanmass}) will again
result in Eq.~(\ref{eq:lean}) but with
$\rho_L=(\rho_P+k\rho_C)/((1+h_p + k +k h_G)$.  For $k=0.044<<1$ as
suggested by Snyder et al. \cite{snyder1984},  $\rho_L$ has approximately the same value as before.

Previous studies have considered two dimensional
models of body mass change although they were
not derived from the three dimensional macronutrient partition model.
Alpert~\cite{alpert1979,alpert1990,alpert2005} considered a model
with $E$ linearized in $F$ 
and $L$ and 
different $f$ depending on context.  
Forbes~\cite{Forbes1970} and Livingston~et al.~\cite{Livingston2001}
modeled weight loss as a 
double exponential decay.  Although, they did not consider
macronutrient flux balance, the dynamics of their models
are equivalent
to the two dimensional model with $I_F$ and $I_L$ zero, and
$E$ linear in $F$ and $L$.

\subsubsection{Energy partition model}\label{energy}

The two compartment macronutrient partition model can be further
simplified by assuming that
trajectories in 
the $L$ - $F$ phase plane 
follow prescribed paths satisfying
\begin{equation}
 \frac{\rho_F}{\rho_L}\frac{dF}{dL}=\alpha(F,L)
 \label{eq:single}
\end{equation}
where $\alpha(F,L)$ is a continuous
function~\cite{Forbes1987,hbc,jordan}
that depends on the mechanisms of body weight
change. This stringent constraint was first hypothesized
by Forbes after analyzing body composition data collected
across a large number of 
subjects \cite{Forbes1987,Forbes2000}.
Forbes postulated that for adults
\begin{equation}
\alpha=\frac{\rho_F}{\rho_L}\frac{F}{10.4}
\label{forbes}
\end{equation}
so that
\begin{equation}
F=D\exp(L/10.4)
\label{forbeslaw}
\end{equation}
where $D$ is a free parameter, and the lean and fat masses are
in units of kg.
Forbes found that his general relationship (\ref{forbeslaw}) is
similar whether weight 
loss is induced by diet or exercise \cite{Forbes2000}. 
It is possible that resistance exercise or a significant
change in the protein content of the diet may result in a different
relationship for $\alpha$
\cite{hansen2007,layman2005,stiegler2006}. Infant growth is 
an example where $\alpha$ is not well described by the Forbes
relationship. Jordan and Hall \cite{jordan} used longitudinal body
composition 
data in growing infants to determine an appropriate form for $\alpha$
during the first two years of life.

Equation
(\ref{eq:single}) describes a family of  $F$ vs $L$ curves,
parameterized by an integration constant (e.g. D in
Eq. (\ref{forbeslaw})).  Depending 
on the initial condition,  the body composition moves along one of
these curves when out of energy balance. Dividing Eq.~(\ref{eq:fat}) by
Eq.~(\ref{eq:lean}) and imposing 
Eq.~(\ref{eq:single})  results in
\begin{equation}
%f(F,L)=\frac{I_F}{E}-\frac{\alpha}{1+\alpha}\frac{I-E}{E}
f(F,L)=\frac{I_F-\alpha I_L+\alpha E}{(1+\alpha)E}=\frac{I_F}{E}-\frac{\alpha}{1+\alpha}\frac{I-E}{E}
\label{f}
\end{equation}
Hall, Bain
and Chow~\cite{hbc} showed that the two compartment macronutrient
partition model 
with  Eq.~(\ref{f}) using Forbes's law~(\ref{forbes}) 
matched a wide range  
of data without any adjustable parameters.

Substituting Eq.~(\ref{f}) into the macronutrient partition model
(\ref{eq:fat}) and (\ref{eq:lean}) 
leads to the {\it Energy Partition} model:  
\begin{eqnarray}
\rho_F\frac{dF}{dt}&=&(1-p)(I-E)\label{eq:fat1}\\
\rho_L\frac{dL}{dt}&=&p(I-E)\label{eq:lean1}
\end{eqnarray}
where $p=p(F,L)=1/(1+\alpha)$ is known as the p-ratio~\cite{dugdale1977}.  
In the energy partition model,  an energy
imbalance $I-E$ is divided between the compartments according to a function
$p(F,L)$ that defines the fraction assigned to lean body tissue
(mostly protein).  Most of the previous models in the literature are
different versions of the 
energy partition 
model~\cite{antonetti1973,payne1977b,alpert1990,Girardier1994,westerterp1995,kozusko2001,flatt2004,christiansen2005},
although none of the authors have noted the connection to macronutrient
flux balance 
or analyzed their models using dynamical systems theory.   Some of these previous models are expressed as computational algorithms that can be translated to the form of the energy partition model.

Despite the ubiquity of the energy partition model, the physiological 
interpretation of the p-ratio remains obscure and is difficult to
measure directly.  It can be inferred indirectly from $f$ (which can
be measured by indirect calorimetry) by using
Eq.~(\ref{f})~\cite{hbc}.  
Previous uses of
the energy partition model often considered $p$ to be a
constant~\cite{antonetti1973,payne1977b,alpert1990,Girardier1994,kozusko2001,flatt2004,christiansen2005}, which implies that the partitioning of energy is independent of
current body composition and macronutrient composition.  This is in
contradiction to 
weight loss data that finds that the fraction of body fat lost does
depend on body 
composition with more fat lost if the body fat is initially
higher~\cite{Forbes1987,elia1999,hall2007b}.
However, if $\alpha$ is a weak function of body composition then a constant
p-ratio may be a valid approximation for small changes. 
Flatt~\cite{flatt2004} considered a model where the p-ratio was
constant but included
the dynamics for glycogen.  His model would be useful when dynamics on
short time scales are of interest.

It  may sometimes be convenient to express the  macronutrient partition model with a unique fixed point as
\begin{eqnarray}
\rho_F\frac{dF}{dt}&=&(1-p)(I-E)+\psi\label{altf}\\
\rho_L\frac{dL}{dt}&=&p(I-E)-\psi
\label{altl}
\end{eqnarray}
for a function $\psi=\psi(I_F,I_L,F,L)$, which is zero at the
fixed point $(F_0,L_0)$.   We use this form in numerical examples in Sec.~\ref{numsim}.
The fasting model of Song and Thomas
\cite{song2007}) used this form with $I=0$ and $\psi$ was a function of $F$
representing ketone production.  
 Comparing to Eq.~(\ref{eq:fat}) and Eq.~(\ref{altf}) gives
\begin{equation}
f=\frac{I_F}{E}-(1-p)\frac{I-E}{E}-\frac{\psi}{E}
\end{equation}

\subsubsection{One Dimensional models}

The dynamics of the energy partition model  Eqs.~(\ref{eq:fat1}) and
(\ref{eq:lean1}) move
along fixed trajectories in the $L$ - $F$ plane.  Thus a further
simplification to a one dimensional model is possible by finding
a functional relationship between
$F$ and $L$ so that one variable can be eliminated in
favor of the other.  
Such a function exists if Eq.~(\ref{eq:single}) has
a unique solution, which is guaranteed in some interval of $L$ if
$\alpha(F,L)$ and $\partial \alpha/\partial F$ are continuous
functions of $F$ and $L$ on a rectangle containing this interval.
These are 
sufficient but not necessary conditions. 

Suppose a relationship $F=\phi(L)$ can be found between $F$ and $L$.
Substituting this relationship into Eq.~(\ref{eq:fat1})
and Eq.~(\ref{eq:lean1}) and adding the two resulting equations yields the
one dimensional equation 
\begin{equation}
\frac{dL}{dt}=\frac{I-E(\phi(L),L)}{\rho_F\phi'(L)+\rho_L}
\label{onedL}
\end{equation}
We can obtain a dynamical equation for body mass by expressing the
body mass as $M=L+\phi(L)$.  If we can invert this relationship uniquely and
obtain $L$  as a function of $M$, then this  can be substituted into
Eq.~(\ref{onedL}) to obtain a dynamical equation for $M$.

As an example, assume $p$ to be a constant, which was used in~\cite{antonetti1973,payne1977b,alpert1990,Girardier1994,kozusko2001,flatt2004,christiansen2005}.
This implies that the phase orbits are a family of straight lines of the form
$F=\beta L+C\equiv \phi(L)$ where $\beta=\rho_L(1-p)/(\rho_Fp)$ and
$C$ is a constant that is specified by the initial body
composition.  This results in
\begin{equation}
\frac{dM}{dt}=\left(\frac{1-p}{\rho_F}+\frac{p}{\rho_L}\right)
\left[I-E\left(\frac{\beta}{1+\beta}(M-C)+C,\frac{M-C}{1+\beta}\right)\right] 
\label{eq:oned1}
\end{equation}
Linearizing Eq.~(\ref{eq:oned1}) around a mass $M_0$ gives
\begin{equation}
\rho_M\frac{dM}{dt}= \mu - \epsilon (M-M_0)
\label{eq:1d}
\end{equation}
where  $\rho_M=\rho_F\rho_L/(\rho_L+(\rho_F-\rho_L)p)$,
$\mu=I-E(F(M_0),L(M_0))$ and $\epsilon=dE/dM|_{M=M_0}$.  This is the form used 
in \cite{christiansen2005}.

If Eqs.~(\ref{eq:fat1}) and (\ref{eq:lean1}) are constrained to
obey the phase plane paths 
of Forbes's law, then a reduction to a one
dimensional equation can also be made. Using Eq.~(\ref{forbeslaw})
(i.e. $\phi(L)=D\exp(L/10.4)$)
in Eq.~(\ref{onedL}) yields
\begin{equation}
\frac{dL}{dt}=\frac{10.4}{\rho_FD\exp(L/10.4)+10.4\rho_L}[I-E(D\exp(L/10.4),L)]
\label{eq:l1d}
\end{equation}
Similarly, a one dimensional equation for the fat mass has the form
\begin{equation}
\frac{dF}{dt}=\frac{F}{\rho_FF+10.4\rho_L}[I-E(F,10.4\log(F/D))]
\label{eq:f1d}
\end{equation}
Since the mass functions $M=L+D\exp(L/10.4)$ or $M=F+10.4\log(F/D)$
cannot be inverted in closed form, an explicit one dimensional
differential equation 
in terms of the mass 
cannot be derived.  However, the dynamics of the mass is easily
 obtained using either Eq.~(\ref{eq:l1d}) or
Eq.~(\ref{eq:f1d}) together with the relevant mass function.   For large
changes in body composition, the dynamics could differ significantly
from the constant $p$ models (\ref{eq:oned1}) or (\ref{eq:1d}).

The one dimensional model gives the dynamics of the energy
partition model  along a fixed trajectory in the $F$ - $L$ plane.  The
initial body composition specifies the constant $C$ or $D$ in the above
equations.  A one dimensional model will represent the energy partition
model even if the intake rate is time dependent.  Only for a perturbation
that directly alters body composition will the one dimensional model
no longer apply.  However, after the perturbation
ceases, the one dimensional model with a new constant will apply again.

\subsection{Existence and stability of body weight fixed points}

The various flux balance models can be analyzed using the methods of
dynamical systems theory, which aims to understand dynamics in terms
of the geometric structure of possible trajectories (time courses
of the body components).  If  the models are
smooth and continuous then the global dynamics can be inferred from the
local dynamics of the model near fixed points
(i.e. where the time derivatives of the variables are zero).
To simplify the analysis, we consider the intake
rates to be clamped to constant values or set to
predetermined functions of time.  We do not consider the control and
variation of food intake rate that may arise due to  feedback from the body
composition or from exogenous influences.  
We focus only on what happens to the food once
it is ingested, which is a problem independent of the
control of intake.  We also assume that the averaged energy expenditure rate
does not depend on time explicitly.  Hence, we do not account
for the effects of development, aging or gradual changes in lifestyle,
which 
could lead to an explicit slow time dependence of 
energy expenditure rate.  Thus, our ensuing analysis is mainly applicable to
understanding the slow dynamics of body mass and 
composition for clamped food intake and physical activity over a time
course of months to a few years.

Dynamics in two dimensions are particularly simple to analyze and
can be easily visualized geometrically~\cite{strogatz,gh}.  The one dimensional models
are a subclass of two dimensional dynamics.  Three dimensional
dynamical systems are generally more difficult
to analyze but Hall~\cite{hall2006} found in simulations that the
glycogen levels varied over a small interval and averaged to an
approximate constant for time periods longer than a few days,
implying that the slow dynamics could be effectively captured by a two
dimensional model.  Reduction to
fewer dimensions is an oft used strategy in dynamical systems theory.
Hence, we focus our analysis on two dimensional dynamics.

In two dimensions, changes of body composition and mass are
represented by trajectories 
in the $L$ - $F$ phase plane. 
For $I_F$ and $I_L$ constant, the flux balance model is a two
dimensional autonomous system of ordinary differential equations and
trajectories will flow to attractors.
The only possible attractors are infinity, stable fixed points or stable limit
cycles~\cite{strogatz,gh}.  We note that fixed points within the context of the model
correspond to states of flux balance.
The two compartment macronutrient
partition model is completely general in that all possible autonomous
dynamics in the two dimensional phase plane are realizable.  Any
two or one dimensional autonomous model of body
composition change can be expressed in terms of the two dimensional
macronutrient partition model.

Physical viability constrains $L$ and $F$ to be positive and finite.
For 
differentiable $f$ and $E$, the possible trajectories for fixed
intake rates are completely specified by the dynamics near fixed points of the
system.  Geometrically, the fixed points are given by the
intersections of the nullclines in the $L$ - $F$ plane, which are given
by the solutions of $I_F-fE=0$ and $I_L-(1-f)E=0$. 
Example nullclines and phase plane portraits of the macronutrient
model are shown in 
Fig.~1.   If the nullclines intersect once then there will be a single
fixed point and if it is stable then the steady state body composition and
mass are uniquely determined. 
Multiple intersections can yield multiple stable fixed points implying that
body composition is not unique~\cite{hall2002}.  If the nullclines are
collinear then there can be an attracting one-dimensional invariant
manifold (continuous curve of fixed points) in the $L$ - $F$ plane.  In this 
case, there are an infinite number of possible body compositions for
a fixed diet.  As we will show,
the energy partition model implicitly assumes an invariant manifold.
If a single fixed point exists but is unstable then a stable limit
cycle may exist around it. 
\begin{figure}
\hspace{-.7cm}
	\scalebox{0.6}{\includegraphics{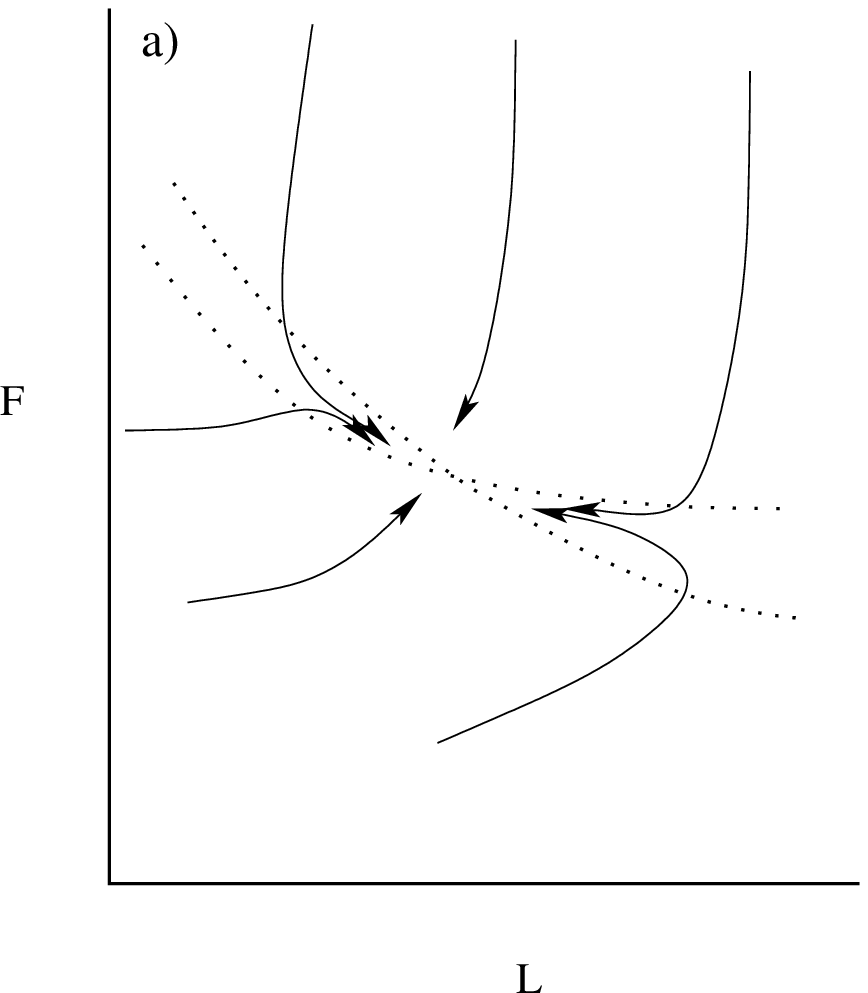}} %\hspace{.1cm}
	\scalebox{0.6}{\includegraphics{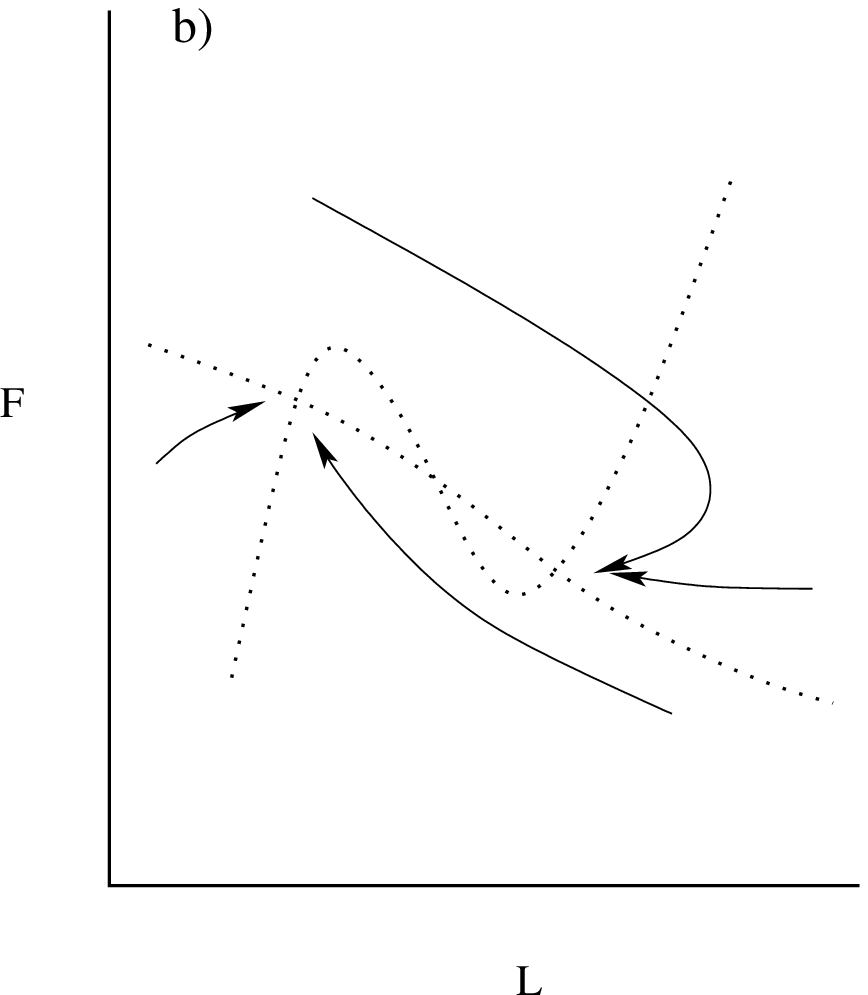}} \\%\hspace{.2cm}
	\scalebox{0.6}{\includegraphics{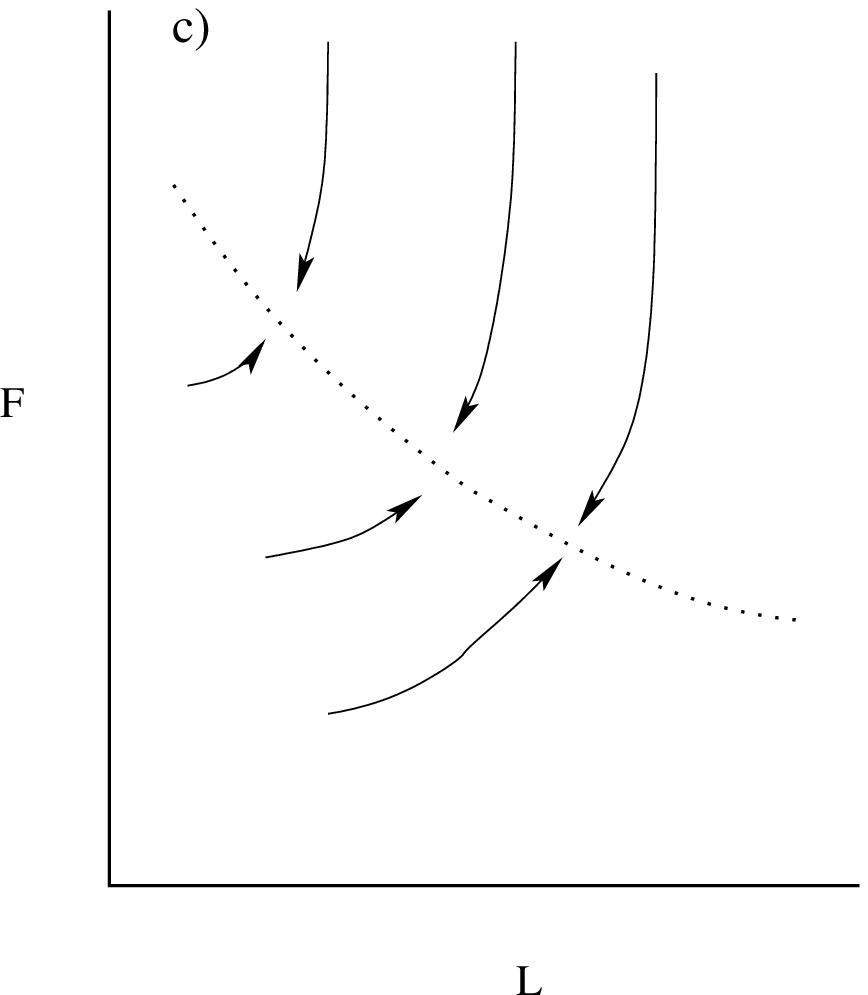}} %\hspace*{.5cm}
	\scalebox{0.6}{\includegraphics{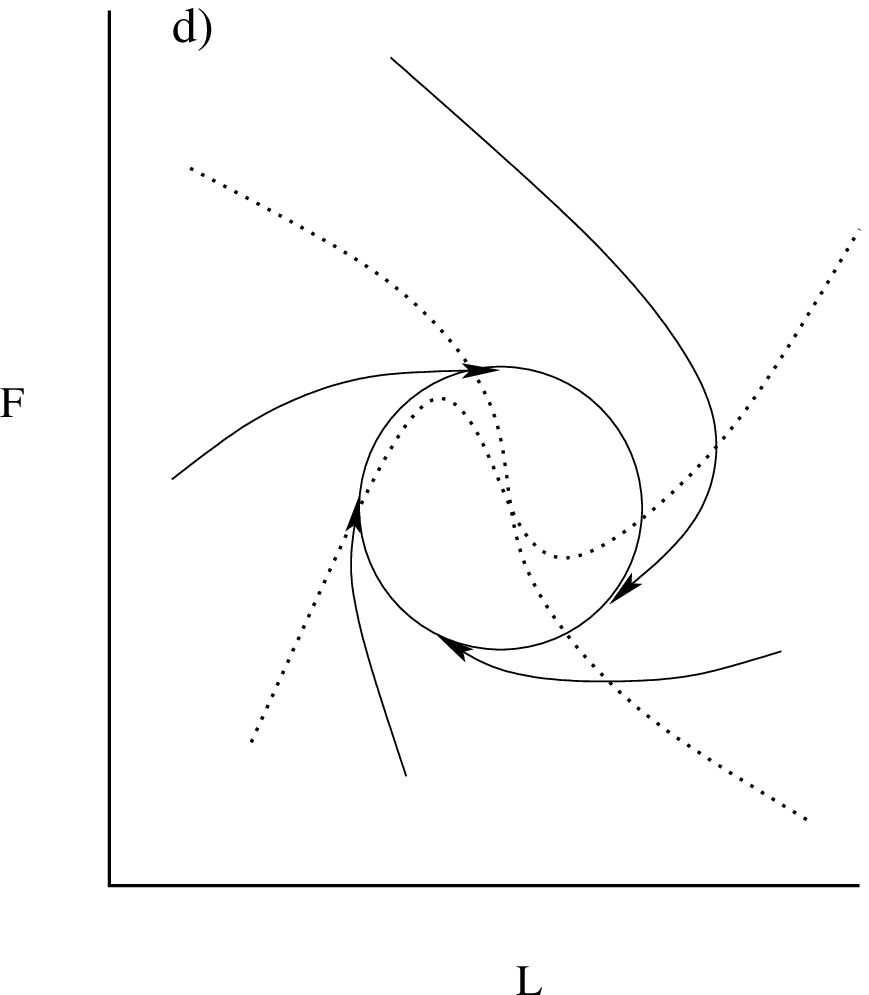}}
	\caption{Possible trajectories (solid lines) for different
	initial conditions and nullclines
	(dotted lines) in the $L$ - $F$ phase plane for models with
	a stable fixed point a), multi-stability with two stable
	fixed points separated by one unstable saddle 
	point  b), an attracting invariant
	manifold c), and a limit cycle
	attractor d).} 
	\label{pp}
\end{figure}
\\

The fixed point conditions of Eqs.~(\ref{eq:fat}) and (\ref{eq:lean}) can
be expressed in terms of
the solutions of 
\begin{eqnarray}
E(F,L)=I\label{eq:IE}\\
f(F,L)=\frac{I_F}{I}\label{eq:feq}
\end{eqnarray}
where $I=I_F+I_L$, and we have suppressed the functional dependence on
intake rates.  These fixed point conditions correspond to a
state of flux balance of the lean and fat components.
Equation~(\ref{eq:IE}) indicates a state of energy balance while
Eq.~(\ref{eq:feq}) indicates that the fraction
of fat utilized must equal the fraction of fat in the diet.
Stability of a fixed point is
determined by the dynamics of small perturbations of body composition
away from the fixed point.  If the perturbed body composition returns
to the original fixed point then the fixed point is deemed stable.  We
give the stability conditions in Methods. 

The functional dependence of $E$ and $f$ on $F$ and $L$ determine the
existence and stability of fixed points.
%Although energy expenditure rate $E$ and fat utilization fraction $f$ are
%measurable quantities, measurements over a 
%wide range of $F$ and $L$ in individual subjects are rare.
As shown in Methods, an isolated stable fixed point is guaranteed if
$f$ is a monotonic  increasing function of $F$
and a monotonic decreasing function of  $L$.  
If one of the fixed
point conditions automatically satisfies the other, then instead of a
fixed point there will be a continuous curve of fixed points or an
invariant manifold.  For example, if the energy balance
condition~(\ref{eq:IE}) automatically satisfies the fat fraction
condition~(\ref{eq:feq}), then there is an invariant manifold 
defined by $I=E$.  The energy
partition model has this property and thus has an invariant manifold
rather than an isolated fixed point.  This can be seen by observing that
for $f$ given by Eq.~(\ref{f}),  $I=E$ automatically satisfies
condition~(\ref{eq:feq}). An attracting
invariant manifold implies that the body 
can exist at any of the infinite number of body compositions specified by the curve $I=E(F,L)$ for clamped intake and energy expenditure rates (see Figure 1 c)).  Each of these infinite possible body compositions will result in a different body mass $M=F+L$ (except for the unlikely case that $E$ is a function of the sum $F+L$).
The body composition
is marginally stable along the direction of the invariant
manifold.  This means that in flux balance, the body composition will
remain at rest at any point on the invariant manifold.  A transient perturbation
along the invariant manifold will simply cause the body composition
to move to a new position on the invariant manifold.
The one dimensional models have a stable fixed point if the invariant
manifold is attracting.
We also show in
Methods that for multiple stable fixed points or a limit cycle to
exist,  $f$ must be nonmonotonic in $L$ and be finely tuned.  The
required fine tuning makes these latter
two possibilities much less plausible than a single fixed point or an
invariant manifold.

Data suggest that $E$ is
a monotonically increasing function of $F$ and
$L$~\cite{cunningham1991}.  The dependence of 
$f$ on $F$ and $L$ is not well established and the form of $f$ depends on
multiple interrelated factors. In general, 
the sensitivity of various tissues to the changing hormonal milieu
will have an overall effect on both the supply of macronutrients as
well as the substrate preferences of various metabolically active
tissues. On the supply side, we know that free fatty acids derived
from adipose tissue lipolysis increase with increasing body fat mass
which thereby increase the daily fat oxidation fraction, $f$, as $F$
increases \cite{astrup1994}. Furthermore, reduction of $F$ with weight loss
has been demonstrated to decrease $f$ \cite{astrup1992}. Similarly,
whole-body proteolysis and protein oxidation increases with lean body
mass \cite{welle1990,short2004} implying that $f$ should be a decreasing
function of $L$. In further support of this relationship, body builders
with significantly increased $L$ have a decreased daily fat oxidation
fraction versus control subjects with similar $F$ \cite{ bosselaers1994}. 
Thus a stable isolated fixed point is consistent with this set of data.

\subsection{Implications for body mass and composition change}\label{implications}

We have shown that all two dimensional autonomous models of  body composition
change generically fall into
two classes -
those with fixed points and those with invariant manifolds. 
In the case of a stable fixed point, any temporary perturbation of
body weight or composition will be corrected over time (i.e.~for all
things equal, the body will return to its original state).  An
invariant manifold allows the possibility that a transient 
perturbation could lead to a permanent change of body composition and
mass.   

At first glance, these differing properties would appear
to point to a simple way of distinguishing between the two
classes. However, the traditional means of inducing weight
change, namely diet or altering energy expenditure through aerobic
exercise, turn out  
to be incapable of revealing the distinction.  For an invariant manifold,
any  change of  intake or expenditure rate
will only elicit movement along one of the prescribed $F$ vs $L$
trajectories obeying Eq.~(\ref{eq:single}),  an example being
Forbes's law (\ref{forbeslaw}). As shown in Fig. 2, a change of intake
or energy expenditure rate will change the position of the invariant
manifold.  The body composition that is initially at one point on
the invariant manifold will then flow to a new point on the perturbed
invariant manifold along the trajectory prescribed by (\ref{eq:single}).  If
the intake rate or energy expenditure is then restored to the
original value then the body composition will return along the same
trajectory to the original steady state just as it would in
a fixed point model (see Fig 2 solid curves).  Only a perturbation that moves the body
composition off of the fixed trajectory could distinguish
between the two classes.  In the fixed point case (Fig. 2 a) dashed-dot curve),
the body composition would go to the same steady state following the perturbation to body composition
but for the invariant manifold case (Fig 2 b) dashed-dot curve), it would go to another
steady state.
\begin{figure}
	\scalebox{.65}{\includegraphics{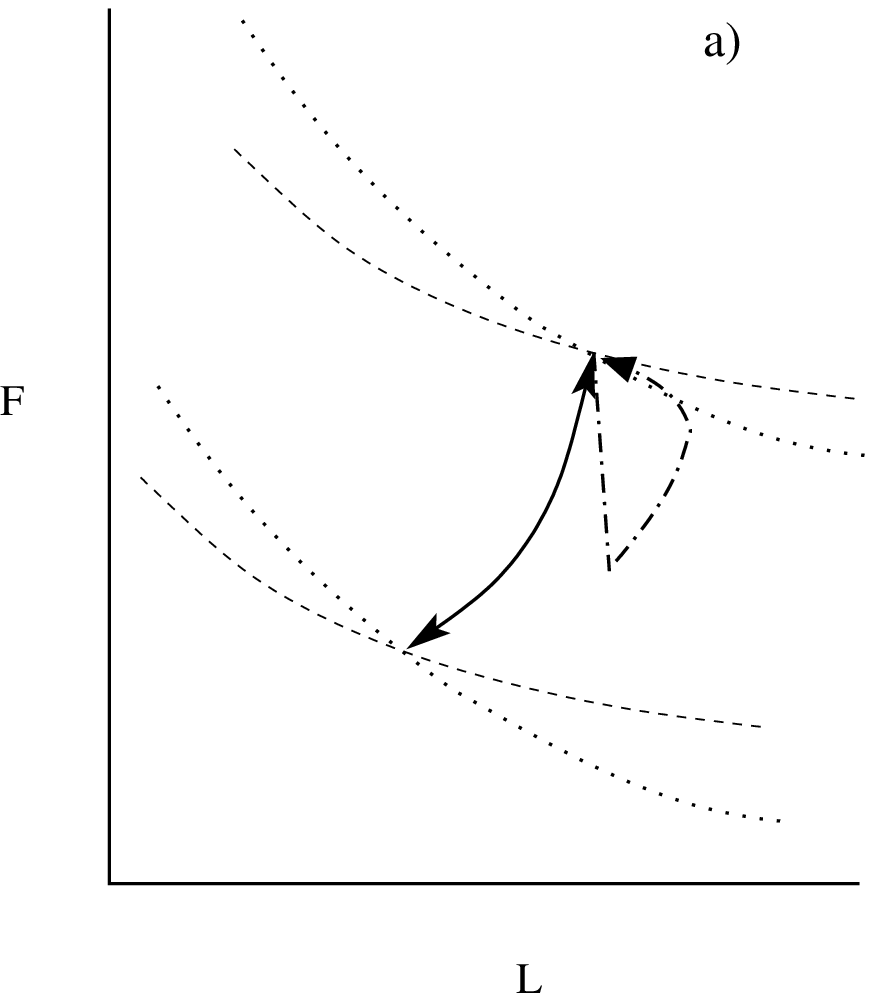}}
	\scalebox{.65}{\includegraphics{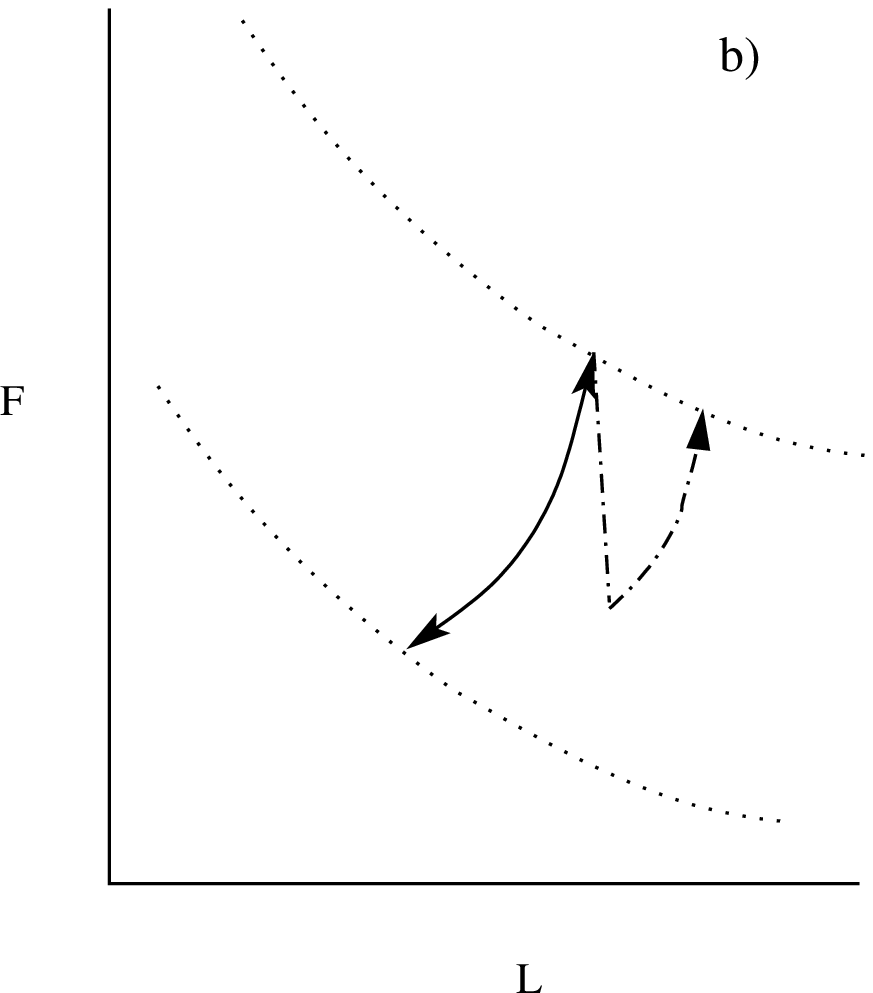}}
	\caption{An example of a situation where the intake or
	energy expenditure rate is changed from one clamped value
	to another and then returned for the fixed 
	point case a) and the invariant manifold case b).  Dotted lines represent nullclines.
	 In both cases, the body composition follows a fixed trajectory and 	returns to
	the original steady state (solid curves).  However, if the body
	composition is perturbed directly (dashed-dot curves) then the body
	composition will flow to same point in a) but to
	a different point in b). 
	}
	\label{fig:invariant}
\end{figure}

Perturbations that move
the body composition off the fixed trajectory can be
done by altering body composition directly or by altering the fat
utilization fraction $f$.
For example, body composition could be altered directly
through liposuction or administering compounds such as growth hormone.
 Resistance exercise may cause an
increase in lean muscle tissue at the 
expense of fat. 
Exogenous hormones, compounds, or infectious
agents that change the propensity for fat versus carbohydrate
oxidation (for example, by increasing adipocyte proliferation and
acting as a sink for fat that is not available for
oxidation \cite{atkinson,pasarica,vangipuram}), would also perturb the
body composition off of a fixed $F$ vs $L$ curve by altering $f$. 
If the body composition returned to its original state after such a
perturbation then there is a unique fixed point.  If it does not then
there could be an invariant manifold although multiple fixed points are also possible.

We found an example of one clinical study that bears on the question of whether humans have a fixed point or an invariant manifold. Biller et al. investigated changes of body composition pre- and post- growth hormone therapy in forty male subjects with growth hormone deficiency \cite{ biller2000}. Despite significant changes of body composition induced by 18 months of growth hormone administration, the subjects returned very closely to their original body composition 18 months following the removal of therapy. However, there was a slight (2\%) but significant increase in their lean body mass compared with the original value.  Perhaps not enough time had elapsed for the lean mass to return to the original level.   Alternatively,  the increased lean mass may possibly have been the result of increased bone mineral mass and extracellular fluid expansion, both of which are known effects of growth hormone, but were assumed to be constant in the body composition models.
Therefore, this clinical study provides some evidence in support of a fixed point, but it has not been repeated and the result was not conclusive.
Using data from the Minnesota
experiment \cite{keys1950} and the underlying physiology,
Hall~\cite{hall2006} proposed a form for $f$ that predicts a fixed point.      On the other hand, Hall, Bain and Chow~\cite{hbc} showed that an invariant manifold
 model is consistent with existing data of
 longitudinal weight change but these experiments only altered
 weight through changes in caloric intake so this cannot rule out the possibility of a
 fixed point.  Thus it appears that existing data is insufficient to decide the
 issue.

\subsubsection{Numerical simulations}\label{numsim}

We now consider some numerical examples using
the macronutrient partition model in the form given by Eqs.~(\ref{altf}) and
(\ref{altl}), with a p-ratio consistent with Forbes's law
(\ref{forbes}): $p=2/(2+F)$, where $F$ is in units of kg.  Consider two cases of the model. 
If $\psi=0$ then 
the model has an invariant manifold and body composition moves along a fixed trajectory in the $L$ - $F$ plane.  If $\psi$ is nonzero, then there
can be an isolated fixed point.  We will show an example where if the intake energy is perturbed, the approach of the body composition to the steady state will be identical for both cases but if body composition is perturbed, the body will arrive at different steady states.

For every model with an invariant manifold, a model with a fixed point can be found such that trajectories in the \mbox{$L$ - $F$} plane resulting from energy intake perturbations will be identical.  All that is required is that $\psi$ in the fixed point model is chosen such that the solution of $\psi(F,L)=0$ defines the fixed trajectory of the invariant manifold model.  Using Forbes's law (\ref{forbeslaw}), we choose $\psi= 0.05(F-0.4\exp(L/10.4))/F$.
We then take a plausible
energy expenditure rate of $E=0.14 L + 0.05 F +1.55$, where energy
rate has units 
of MJ/day and mass has units of kg.  This expression is based on
combining cross-sectional data~\cite{cunningham1991} for
resting energy with a contribution of physical activity of a fairly
sedentary 
person~\cite{hall2006}. 
Previous models propose similar 
forms for the energy
expenditure~\cite{payne1977b,alpert1979,alpert2005,christiansen2005}.

Figure~3 shows the time dependence of body mass and the $F$ vs
$L$ trajectories of the two model examples given a
reduction in energy intake rate from 12 MJ/day to 10 MJ/day starting at the
same initial condition.  The time courses are identical for body composition and mass.
The mass first decreases linearly in time but then  
saturates to a new stable fixed point. 
The dashed line represents the same intake rate
reduction but with 10 kg of fat removed at day 100.
For
the invariant manifold model, the fat perturbation
permanently alters the final body composition and body mass, whereas in the
fixed point model it only has a transient effect.  
In the fixed point model, the body
composition can ultimately exist only at one point given by the
intersection of the  
nullclines (i.e.~solution of $I=E$ and
$\psi=0$).  For the invariant manifold, the body composition can exist
at any point on the $I=E$ curve (dotted line in Fig.~2 d)). 
Since a $\psi$ can always be found so that a fixed point model and an
invariant manifold model have identical time courses for body composition and mass, a perturbation
in energy intake can never discriminate between the two possibilities.

The time constant to reach the new fixed point in the numerical simulations is very
long.  This slow approach to steady state (on the order of several years for humans)  has
been pointed out many times
previously~\cite{hall2006,payne1977b,alpert1979,alpert2005,christiansen2005}.
A long time constant will make experiments to distinguish between a
fixed point and an invariant manifold difficult to
conduct.   Experimentally reproducing this example would be
demanding but if the time variation of the intake rates and
physical activity levels were small compared to the induced change
then the same result  should arise qualitatively.  Additionally,
the time
constant depends on the form of the energy expenditure.
There is evidence that the dependence of energy expenditure on $F$ and
$L$ for an individual is steeper than for the population due to an
effect called adaptive thermogenesis~\cite{leibel}, thus making
the time constant shorter.
\begin{figure}
\hspace{-.7cm}
	\scalebox{.65}{\includegraphics{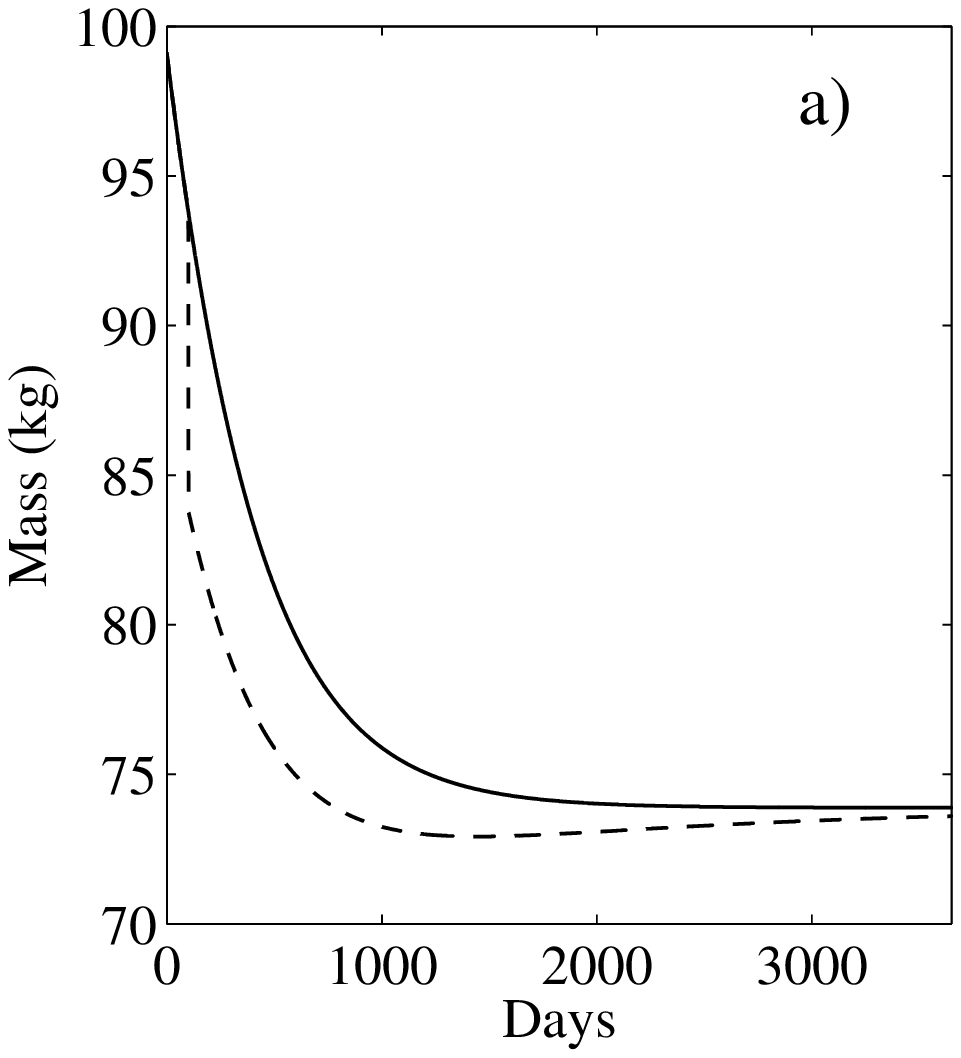}}
	\scalebox{.68}{\includegraphics{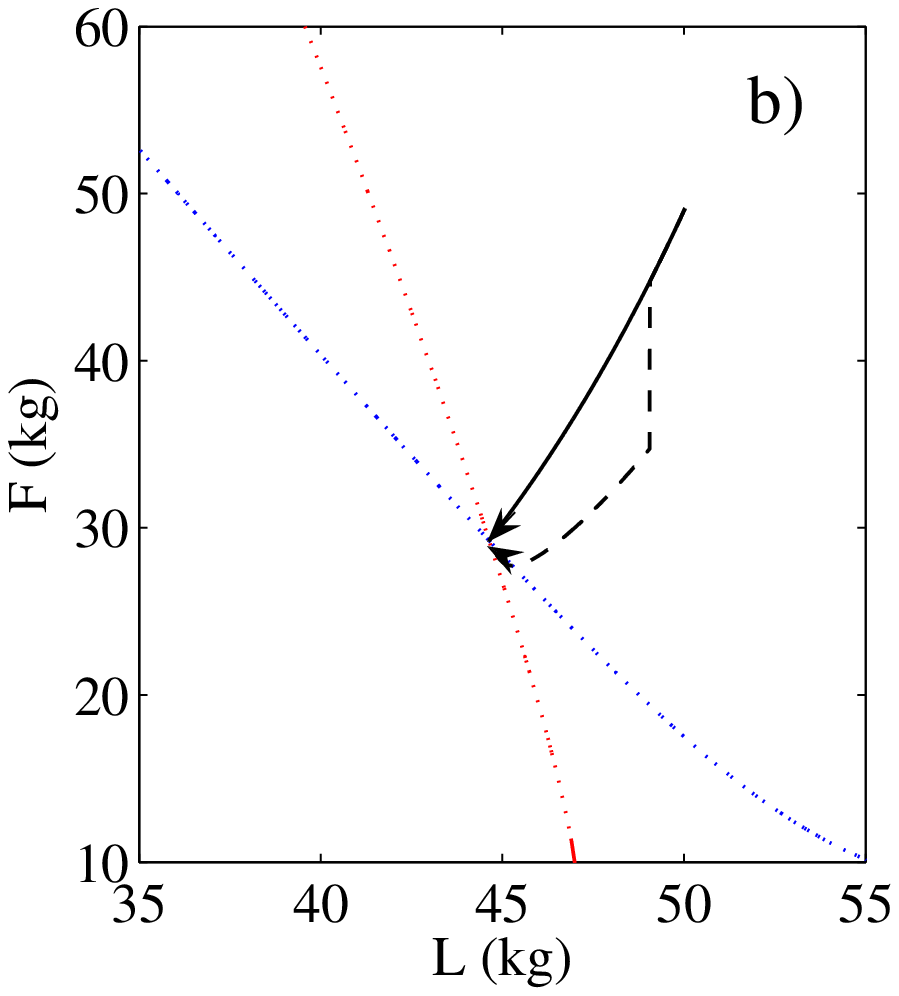}}
	\scalebox{.65}{\includegraphics{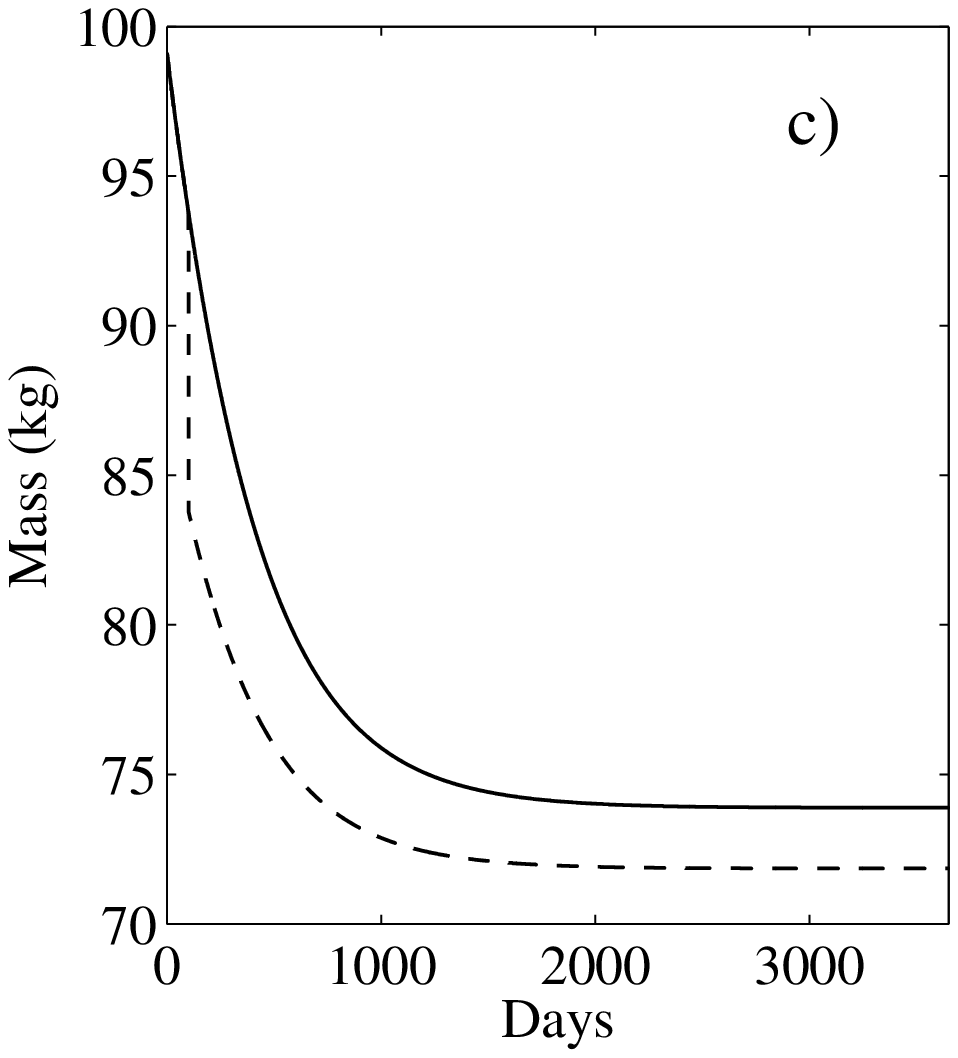}}
\hspace{-.2cm}
	\scalebox{.68}{\includegraphics{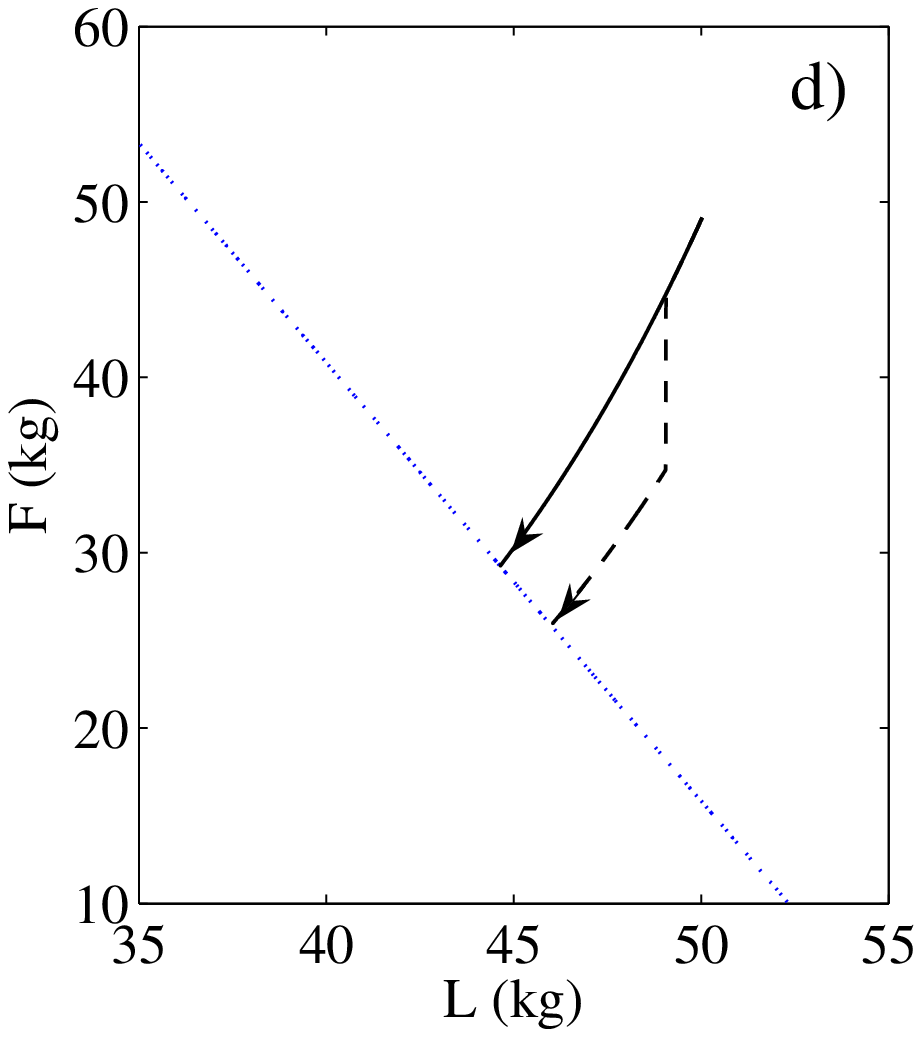}}
	\caption{Time dependence of body mass for fixed point
	model a). In all the figures, the solid line is for an intake reduction
	from 12 MJ/day to 10 J/day and the dashed line is for the same
	reduction but with a removal of 10 kg of fat at day 100.
	Trajectories in the $F$ vs $L$ phase plane for the fixed
	point model b).  Dotted lines are the nullclines.  Time
	dependence c) and phase plane d) of the invariant manifold
	model for the same conditions.
	}
	\label{fig:fp}
\end{figure}

\section{Discussion}

In this paper we have shown that all possible two dimensional
autonomous models
for lean and fat mass are variants of the macronutrient partition
model.  The models can be divided into two general classes - models
with
isolated fixed points (most likely a single stable fixed point) and
models with an invariant manifold.  There 
is the possibility of more exotic behavior such as multi-stability  and
limit cycles but these require fine tuning and thus are less plausible. 
Surprisingly, experimentally  
determining if the body exhibits a fixed point or an invariant
manifold is nontrivial.
Only perturbations  
of the body composition itself apart from dietary or energy
expenditure interventions or alterations of the fraction of energy
utilized as fat can discriminate between the two possibilities.
The distinction between the classes is not merely an academic concern
since this has direct clinical implications for potential permanence
of transient changes of body composition via such procedures as
liposuction or temporary  administration of therapeutic compounds.  

Our analysis considers the slow dynamics of the body mass and  composition
where the fast time dependent hourly or daily fluctuations are averaged
out for a clamped average food intake rate. We also do not
consider a slow explicit time dependence of the energy expenditure.
Such a time dependence could arise during development, aging or gradual
changes in lifestyle where activity levels differ.  Thus our analysis
is best suited to modeling changes over time scales of months to a few
years in adults.  We do not consider any feedback of body composition
on food intake, which is an extremely important topic  but beyond
the scope of this paper.

Previous efforts to model body weight change have predominantly used
energy partition models that implicitly contain an
invariant manifold and thus body composition and mass are not 
fully specified by the diet.  If the body does have an invariant
manifold then this fact puts a very strong constraint on the fat
utilization fraction $f$.  Hall~\cite{hall2006} considered the effects of
carbohydrate intake on lipolysis and other physiological factors to
conjecture a 
form of $f$ that does not lead to an invariant manifold.  However, our
analysis 
and numerical examples show that the body composition could have an invariant
manifold
but
behave indistinguishably from having a fixed point. Also, the
decay to the fixed point could take a very long
time, possibly as long as a decade giving the appearance of an
invariant manifold. Only experiments that perturb the fat or lean
compartments independently 
can tell.

\section{Methods}

\subsection{Method of Averaging}

The three compartment macronutrient flux balance equations
(\ref{eq:ffat})-(\ref{eq:prot}) are a
system of nonautonomous
differential equations since the energy intake and expenditure
are explicitly time dependent.  Food is ingested over discrete time intervals and physical activity will vary greatly within a day.
However, this fast time dependence can be
viewed as oscillations or fluctuations on top of a slowly varying background.
It is this slower time dependence that governs long-term body mass and
composition changes
that we are interested in.  For example, if an individual had the exact same schedule with the same energy intake and expenditure each day, then averaged over a day, the body composition would be constant.  If the daily averaged intake and expenditure were to gradually change on longer time scales of say weeks or months then there would be a corresponding change in the body composition and mass.  Given that we are only interested in these slower changes, we remove the short time scale
fluctuations by using the method of averaging to
produce an autonomous system of {\em averaged} equations valid on longer
time scales. 

We do so by introducing a second ``fast" time variable $\tau=t/\epsilon$, where $\epsilon$ is a small parameter that is associated with the slow changes and let all time
dependent quantities be a function of both $t$ and $\tau$.  For example, if $t$ is measured in units of days and $\tau$ is measured in units of hours then $\epsilon\sim 1/24$.
Inserting into
(\ref{eq:ffat})-(\ref{eq:prot}) and using the chain rule yields
\begin{eqnarray}
\rho_F\left(\frac{\partial F}{\partial t}+\frac{1}{\epsilon}\frac{\partial F}{\partial\tau}\right)&=&I_F(t,\tau)-f_F E(t,\tau)\label{eq:ffat2}\\
\rho_G\left(\frac{\partial G}{\partial t}+\frac{1}{\epsilon}\frac{\partial G}{\partial\tau}\right)&=&I_C(t,\tau)-f_C E(t,\tau)\label{eq:glyc2}\\
\rho_P\left(\frac{\partial P}{\partial t}+\frac{1}{\epsilon}\frac{\partial P}{\partial\tau}\right)&=&I_P(t,\tau)-(1-f_F-f_C) E(t,\tau) \label{eq:prot2}
\end{eqnarray}
We then consider
the three body compartments to have expansions of the form
\begin{eqnarray}
F(t,\tau)&=&F^0(t)+\epsilon F^1(t,\tau)+O(\epsilon^2)\\
G(t,\tau)&=&G^0(t)+\epsilon G^1(t,\tau)+O(\epsilon^2)\\
P(t,\tau)&=&P^0(t)+\epsilon P^1(t,\tau)+O(\epsilon^2)
\end{eqnarray}
where $\langle F^1\rangle=\langle P^1\rangle = \langle G^1\rangle=0$
for a time average defined by
$\langle X
\rangle=(1/T)\int_0^{T} X \  d\tau$ 
and $T$ represents an averaging time scale of a day.   The fast time dependence can be either
periodic or stochastic.  The important thing is that the time average
over the fast quantities is of order $\epsilon$ or higher.
We then expand the energy expenditure rate
and expenditure 
fractions to first order
in $\epsilon$:
\begin{equation}
E(F,G,P,t,\tau)= E^0(t,\tau)+\epsilon\left(\frac{\partial
  E}{\partial F} F^1 + \frac{\partial
  E}{\partial G} G^1 + \frac{\partial
  E}{\partial P} P^1\right)+O(\epsilon^2)
\end{equation}
\begin{eqnarray}
f^0_i(F,G,P)=f_i(F^0,G^0,P^0)+\epsilon\left(\frac{\partial
  f_i}{\partial F} F^1 + \frac{\partial
  f_i}{\partial G} G^1 + \frac{\partial
  f_i}{\partial P} P^1\right) +O(\epsilon^2)
\end{eqnarray}
where $E^0(t,\tau)\equiv E(F^0,G^0,P^0,t,\tau)+O(\epsilon^2)$ and
$i\in \{F,G,P\}$.  We assume that the
expenditure fractions depend on time only
through the body compartments.
Substituting these expansions into
Eqs.~(\ref{eq:ffat2})-(\ref{eq:prot2}) and taking lowest 
order in $\epsilon$ gives
\begin{eqnarray}
\rho_F\left(\frac{\partial F^0}{\partial t}+\frac{\partial F^1}{\partial\tau}\right)&=&I_F(t,\tau)-f^0_F E^0(t,\tau)\label{eq:ffat3}\\
\rho_G\left(\frac{\partial G^0}{\partial t}+\frac{\partial G^1}{\partial\tau}\right)&=&I_C(t,\tau)-f^0_C E^0(t,\tau)\label{eq:glyc3}\\
\rho_P\left(\frac{\partial P^0}{\partial t}+\frac{\partial P^1}{\partial\tau}\right)&=&I_P(t,\tau)-(1-f_F-f^0_C) E^0(t,\tau) \label{eq:prot3}
\end{eqnarray}

Taking the moving time average of
Eqs.~(\ref{eq:ffat3})-(\ref{eq:prot3}) and requiring that $\langle
\partial F^1/\partial \tau \rangle$,  $ \langle
\partial G^1/\partial \tau \rangle$, and $\langle
\partial P^1/\partial \tau \rangle$ are of order $\epsilon$ or higher
leads to the averaged equations:
\begin{eqnarray}
\rho_F\frac{dF^0}{dt}&=&\langle I_F\rangle -f^0_F \langle E^0\rangle\label{eq:ffat4}\\
\rho_G\frac{dG^0}{dt}&=&\langle I_C\rangle -f^0_C \langle E^0\rangle\label{eq:glyc4}\\
\rho_P\frac{dP^0}{dt}&=&\langle I_P\rangle -(1-f^0_F-f^0_C) \langle E^0\rangle \label{eq:prot4}
\end{eqnarray}
In the main text we only consider the slow time scale dynamics so we
drop the superscript and bracket notation for simplicity.  Hence, the
system (\ref{eq:ffat})-(\ref{eq:prot}) can be thought of as
representing the lowest order time averaged macronutrient flux
balance equations.  We note that in addition to the daily fluctuations of meals and physical activity, there can also be fluctuations in food intake from day to day \cite{periwal2006}.  Our averaging scheme can be used to average over these fluctuations as well by extending the averaging time $T$.   A difference
in the choice of $T$ will only result in a different interpretation of
the averaged quantities.

\subsection{Stability conditions for fixed points}

The dynamics near a fixed point $(F_0,L_0)$ are determined by expanding
$fE$ and $(1-f)E$ to linear order in $\delta F=F-F_0$ and $\delta
L=L-L_0$ \cite{strogatz,gh}.  Assuming solutions of the form $\exp(\lambda t)$ yields an
eigenvalue problem with two
eigenvalues given by 
$\lambda=\frac{1}{2}\left({\rm Tr} J \pm \sqrt{{\rm Tr}J^2  -4\det
  J}\right)$
where
\begin{equation}
{\rm Tr} J=  
-\left[\frac{1}{\rho_F}\frac{\partial}{\partial F}(f E)+\frac{1}{\rho_L}\frac{\partial}{\partial L}((1-f) E)\right]_{(F_0,L_0)}
\label{eiv}
\end{equation}
and
\begin{equation}
\det J=
%\frac{E}{\rho_L\rho_F}\left[ \partial_L E\partial_F f-\partial_F E\partial_L f
\frac{E}{\rho_L\rho_F}\left[ \frac{\partial E}{\partial L} \frac{\partial f}{\partial F}-
\frac{\partial E}{\partial F}\frac{\partial f}{\partial L}
\right]_{(F_0,L_0)}
\label{det}
\end{equation}
A fixed point is stable if and only if ${\rm Tr} J<0$ and $\det J>0$. 
In the case of an invariant manifold, $\det J=0$, so the
eigenvalues are ${\rm Tr} J$ and $0$.  The zero eigenvalue reflects
the marginal stability along the invariant manifold, which is an
attractor if ${\rm Tr} J <0$.  An attracting invariant manifold
implies a stable fixed point in the corresponding one dimensional model.
Unstable fixed points are either unstable nodes, saddle points or
unstable spirals.  In the case of unstable spirals, a possibility is a
limit cycle surrounding the spiral arising from a Hopf bifurcation,
where ${\rm Tr}J=0$ and $\det J>0$.   In this case, body composition
and mass would 
oscillate even if the intake rates were held constant. The frequency
and amplitude of the oscillations may be estimated near a
supercritical Hopf
bifurcation by transforming the equations to normal form.  
Stability of a fixed point puts constraints on the form of
$f$.  Physiological considerations and data imply that
$\partial E /\partial L> \partial E/\partial F
>0$~\cite{cunningham1991,hall2006}.  Thus we can set $\partial
E/\partial F=\delta  
\partial E /\partial L$ where $\delta<1$ (where the derivatives are evaluated at the fixed point).  Then $\det J>0$ implies that  
\begin{equation}
\partial f/\partial F> \delta \partial f/\partial L
\label{conddet}
\end{equation}
 and  ${\rm Tr} J < 0$ implies 
\begin{equation}
\partial f/\partial F> \gamma\partial f/\partial L  - K,
\label{condtr}
\end{equation}
where $K=[\delta f+\gamma(1-f)](\partial E/\partial L)/E>0$ and
$\gamma=\rho_F/\rho_L\simeq 5.2$.  Hence $\partial f/\partial F>0$ and
$\partial f/\partial L <0$ guarantees stability of a fixed point.

From Eq.~(\ref{conddet}) 
and Eq.~(\ref{condtr}) and the fixed point conditions (\ref{eq:IE}) and
(\ref{eq:feq}), if $f$ increases monotonically with $F$ and decreases
monotonically with $L$ then there will be a unique stable fixed point.
For an invariant manifold,  $f$ is given by Eq. (\ref{f}), which immediately satisfies $\det J=0$; ${\rm Tr} J<0$ is guaranteed if $E$ is monotonically increasing in $F$ and $L$.  
For a Hopf bifurcation, we require $\partial f/\partial
F=\gamma\partial f/\partial L  - K$ and Eq.~(\ref{conddet}), implying
$(\gamma-\delta)\partial f/\partial L -K>0$.  Since $\gamma>\delta$,
$f$ must increase 
with $L$ for the possibility of a limit cycle.  However, to ensure
that trajectories remain bounded $f$ must decrease with $L$ for very
small and large values of $L$.  Hence, $f$ must be nonmonotonic in $L$
for a limit cycle to exist.  This can also be seen from an application of
Bendixson's criterion~\cite{gh}, which states that a limit cycle cannot exist in a given region of the $L$ - $F$ plane if
\begin{equation}
\frac{1}{\rho_F}\frac{\partial}{\partial F}(f E)+\frac{1}{\rho_L}\frac{\partial}{\partial L}((1-f) E)
\end{equation}
does not change sign in that region.  In addition, the other
parameters must be fine tuned  for a limit cycle (see Fig.~1 d)).   Similarly, as seen
in Fig.~1 c), for multi-stability to exist, nonmonotonicity and fine
tuning are also required.  

\begin{acknowledgments}
This research
was supported by the Intramural Research Program of the NIH, NIDDK.  
\end{acknowledgments}

\bibliographystyle{unsrt}

\bibliography{modelclasses.bib}
%\bibliography{met_refs.bib}

\end{document}